\documentclass[twocolumn,showpacs,amsmath,pra,aps,amssymb,superscriptaddress,nofootinbib]{revtex4-1} 
\usepackage[T1]{fontenc}
\usepackage[latin9]{inputenc}
\usepackage{times}
\usepackage{color} 
\usepackage{xspace}
\usepackage{amssymb,amsmath}
\usepackage{amsbsy}
\usepackage{dsfont}
\usepackage[pdftex]{graphicx}
\usepackage{bm}
\usepackage{float}
\usepackage{adjustbox}
\usepackage{epstopdf}

\usepackage[unicode,breaklinks]{hyperref}
\hypersetup{
    unicode=true,
    plainpages=false, 
    colorlinks=true,
    linkcolor=blue,
    citecolor=blue,
    filecolor=black,
    urlcolor=blue
}
\usepackage{url}
\usepackage{verbatim}

\begin{document}

\title{Solving the spin-2 Gross-Pitaevskii equation using exact nonlinear dynamics and symplectic composition}
\author{L.~M.~Symes}  
\affiliation{Dodd-Walls Centre for Photonic and Quantum Technologies, Department of Physics, University of Otago, Dunedin 9016, New Zealand}
\author{P.~B.~Blakie}  
\affiliation{Dodd-Walls Centre for Photonic and Quantum Technologies, Department of Physics, University of Otago, Dunedin 9016, New Zealand}

\begin{abstract}
We develop numerical methods for solving the spin-2 Gross-Pitaevskii equation. The basis of our work is a  two-way splitting of this evolution equation that leads to two exactly solvable subsystems. Utilizing second-order and fourth-order composition schemes we realize two fully symplectic integration algorithms, the first such algorithms for evolving spin-2 condensates. We demonstrate the accuracy of these algorithms against other methods on application to an exact continuous wave solution that we derive.  
\end{abstract}

\maketitle

\section{Introduction}
A spinor Bose-Einstein condensate is a coherent matter-wave in which the constituent atoms are able to access their spin degrees of freedom \cite{Ho1998a,Ohmi1998a,Stenger1998a}. This type of condensate is described by a multi-component field that evolves according to a Gross-Pitaevskii equation (GPE). 
Here we consider the case of a spin-2 system in which the atoms have five internal spin states conveniently labelled by the $z$-projection quantum number as $m=-2,-1,\ldots,+2$. This system has been realized in experiments with ultra-cold $^{87}$Rb  atoms prepared in the $\mathcal{F}=2$ hyperfine manifold \cite{Schmaljohann2004,Chang2004a,Kuwamoto2004}. In weak magnetic fields the short ranged atomic collisions are rotationally invariant and the nonlinear interactions only depend on three parameters: the spin-independent (i.e.~density) interaction strength $c_0$, the spin-dependent (i.e.~spin-density) interaction strength $c_1$,  and the spin-singlet pair interaction strength $c_2$ (see \cite{Kawaguchi2012R}).  The spin-2 case is of great theoretical interest because the order parameters and topological excitations can have non-trivial symmetries, for instance non-Abelian vortices \cite{Semenoff2007,Kobayashi2009a,Huhtamaki2009}.
 
The nonlinear terms in the spinor GPE have been identified as presenting new challenges for mathematical analysis and numerical simulation  \cite{Bao2013a}, as compared to the relatively well-studied problem of the GPE for scalar condensates. 
There have been a number of proposals for schemes to evolve spin-1 condensates (e.g.~see  Refs.~\cite{Wang2007a,Bao2009a,Gawryluk2015a,Symes2016a}). For this case the condensate is described by a three-component field and the evolution equation only involves density ($c_0$) and spin-density ($c_1$) terms.  
For the spin-2 case Wang has developed a time-splitting Fourier pseudo-spectral method \cite{Wang2011a} which is second order accurate in time. This approach makes a three-way splitting of the evolution equation, notably dividing the interaction terms into parts which are diagonal and non-diagonal in the spin state basis. The dynamical evolution arising from the diagonal part conserves the density and can be exactly solved. However, the non-diagonal part is dealt with by numerically diagonalizing it in spin-space,  allowing an approximate solution. 
However, it is desirable to have some analytic treatment to deal with the spinor interaction terms of the spin-2 system. One recent step in this direction was made by Yepez who proposed using an analytical continuation approach to approximately solve the spin-density evolution term \cite{Yepez2016a}.
 
 In this paper we develop two algorithms for evolving the spin-2 GPE, including all the relevant interaction terms. The distinguishing feature of our approach is that we break the evolution equation into just two parts and provide exact analytic solutions for each part. Using second-order and fourth-order symplectic composition rules to combine the solutions of each part we develop the first fully symplectic techniques for solving the spin-2 GPE.
 
The outline of the paper is as follows. We review the description of a spin-2 condensate and present the GPE in Sec.~\ref{SecFormalism}. In Sec.~\ref{SecSplittings} we briefly review how to formally split the spin-2 GPE into parts, solve these parts in isolation, then compose the solutions to produce an approximate solution to the full problem. We also introduce the second and fourth order composition schemes that we use for our numerical algorithm. In Sec.~\ref{SecSpin2} we demonstrate how to split the spin-2 GPE into two parts. The kinetic energy and quadratic Zeeman energy are identified as the first part, and are treated using standard spectral techniques. We identify the second part as consisting of the trap and three interaction terms.  We deal with each of the interaction terms separately, before presenting an analytic solution to the full nonlinear interaction dynamics. We develop an exact continuous wave solution for the spin-2 GPE in Sec.~\ref{SecNumTest}, and use this to test our numerical algorithms and compare with other methods. Finally, we conclude and discuss the outlook for our work.

\section{Formalism} \label{SecFormalism}
\subsection{Spin-2 Gross-Pitaevskii equation}
We consider a spin-2 condensate  
described by the Hamiltonian~\cite{Kawaguchi2012R}%Ho1998a,Ohmi1998a}
\begin{align}\label{spinH}
H &= H_{\mathrm{sp}} + H_{\mathrm{int}}, \\
H_{\mathrm{sp}} &=\int\!d\bm{x}\ \bm{\psi}^\dagger\!\left(\!-\frac{\hbar^2\nabla^2}{2M}+V-pf_z+qf_z^2\right)\!\bm{\psi}, \\
H_{\mathrm{int}} &= \int\!d\bm{x}\left[\frac{c_0}{2}n^2+\frac{c_1}{2}\left|\bm{F}\right|^2 + \frac{c_2}{2} |A_{00}|^2\right]\!.
\end{align}
Here $\bm{\psi}\equiv (\psi_{2}, \psi_{1},\psi_0,\psi_{-1}, \psi_{-2})^T$ is a five component spinor describing the condensate field in the five spin levels and $p$ and $q$ are, respectively, the linear and quadratic Zeeman shifts arising from the presence of a uniform magnetic field along $z$.  The term $V$ represents a spin-independent trapping potential. The term $c_0 n^2$ describes the density interaction with coupling constant $c_0$, where  
\begin{equation}
n\equiv\bm{\psi}^\dagger\bm{\psi},
\end{equation}
is the total density. 
The term $c_1|\bm{F}|^2$ describes the spin interaction with coupling constant $c_1$, where  
\begin{equation}
\bm{F}\equiv \bm{\psi}^\dagger\bm{f}\bm{\psi},
\end{equation}
 is a vector of spin densities arising from the spin-2 matrices $(f_x,f_y,f_z)\equiv\bm{f}$, which are given in Appendix~\ref{app:spin-matrices}.
The term $c_2 |A_{00}|^2$ describes the spin-singlet interaction with coupling constant $c_2$, where 
\begin{equation}
A_{00} \equiv \bm{\psi}^TA\bm{\psi},
\end{equation}
is the spin-singlet, with the matrix $A$ taking the form 
\begin{equation}
A_{m,m'}=\tfrac{1}{\sqrt{5}}(-1)^{m+1}\delta_{m,-m'}.
\end{equation}

The dynamics of the system are described by the spin-2 Gross-Pitaevskii equation 
\begin{align}\label{spinGPEs}
i\dot{\bm{\psi}} &= \left(-\tfrac{1}{2}\nabla^2+ V+q f_z^2 + c_0 n + c_1 \bm{F}\cdot\bm{f}\right)\bm{\psi} + c_2 A_{00} A \bm{\psi}^*,
\end{align}
where we have adopted dimensionless units of distance $x_0$ and time $t_0=mx_0^2/\hbar$, suitably rescaling the interaction coefficients, and we have removed the linear Zeeman term by moving to a rotating frame with the transformation $\bm{\psi}\rightarrow e^{ipf_zt/\hbar}\bm{\psi}$.
 
 \subsection{Conserved quantities}\label{SecConservQuant}
Under evolution according to the spin-2 GPE the condensate conserves 
energy $E$, given by evaluating Eq.~(\ref{spinH}), the normalization
$N= \int d\bm{x}\, n$ and the total $z$-component of magnetization 
$M_z = \int d\bm{x}\, F_z$.

\section{Splittings and symplectic composition}\label{SecSplittings}
\subsection{Two-way splitting}
We denote the spin-2 GPE as $\dot{\bm{\psi}}=\hat{f}\bm{\psi}$, where $\hat{f}$ denotes the nonlinear evolution operator [c.f.~Eq.~(\ref{spinGPEs})], with formal solution $\bm{\psi}(t)=\exp(\hat{f}t)\bm{\psi}(0)$.  For numerical solution we will break up the operator $\hat{f}$ into two parts as
\begin{align}
\hat{f}=\hat{f}_A+\hat{f}_B.
\end{align}
We will then deal with the individual parts separately
\begin{align}
\dot{\bm{\psi}}=\hat{f}_\mu\bm{\psi},\qquad\mu=A,B,
\end{align}
 and  for each we will find an exact solution allowing us to efficiently construct the \emph{flows}
\begin{align}
\bm{\psi}(t)=e^{\hat{f}_\mu t}\bm{\psi}(0),\qquad\mu=A,B.\label{fAB}
\end{align}
\subsection{Symplectic composition schemes}\label{SecSymplectic}
We can approximate the full solution by composing the two subsystem solutions. Since our system is Hamiltonian, it is desirable to use a symplectic composition method in order to maintain the Hamiltonian geometric structure of phase space in our solution.
Here we will consider two different composition schemes \cite{mclachlan2006geometric}. First we use the \textit{Leapfrog} composition as a simple second-order method for advancing the wavefunction by a time step of size $\tau$:
\begin{align}
\bm{\psi}(\tau)=e^{\hat{f}\tau}\bm{\psi}(0)\approx e^{ \hat{f}_A\frac{\tau}{2}} e^{  \hat{f}_B\tau} e^{ \hat{f}_A\frac{\tau}{2}}\bm{\psi}(0).\label{EqLF}
\end{align}
We also consider the fourth-order symplectic composition developed by Blanes and Moan \cite{Blanes2002a,mclachlan2006geometric,Thalhammer2012a}, which takes the form
\begin{align}
\bm{\psi}(\tau)&\approx e^{ \hat{f}_A a_7 \tau} e^{ \hat{f}_B b_6 \tau} e^{ \hat{f}_A a_6 \tau} \cdots e^{ \hat{f}_B b_1 \tau} e^{\hat{f}_A a_1 \tau}\bm{\psi}(0),\label{EqBM}
\end{align}
where the values of the composition coefficients 
are given in Table~2 of \cite{Blanes2002a} and by the relations $a_{8-j}=a_j$ and $b_{7-j}=b_j$.

\section{Splitting of the Spin-2 GPE}\label{SecSpin2}
In this section we discuss the details of the two subsystems and their exact solution.
\subsection{Kinetic and quadratic Zeeman subsystem ($\hat{f}_A$)} \label{secfA}
We take the first subsystem to include the kinetic energy and quadratic Zeeman term, i.e.~we define $\dot{\bm{\psi}}=\hat{f}_A\bm{\psi}$ to be
\begin{align}
\dot{\psi}_m=-i\left[-\tfrac{1}{2}\nabla^2+qm^2\right] \psi_m.\label{IP}
\end{align}
This is diagonal in Fourier space, yielding the solution
\begin{align}
\psi_m(t) &= \mathcal{F}^{-1}\left\{ e^{-i(\frac{1}{2}k^2 + qm^2)t} \mathcal{F} [\psi_m(0)] \right\},
\end{align} 
where $\mathcal{F}$ denotes a Fourier transform of the appropriate spatial dimension for the problem and $k=|\bm{k}|$, with $\bm{k}$ the Fourier space coordinate.

\subsection{Trap and nonlinear spin interactions  ($\hat{f}_B$)} \label{secfB} 

We take the second subsystem to include the trap potential and remaining interaction terms, i.e.~we define $\dot{\bm{\psi}}=\hat{f}_B\bm{\psi}$ to be
\begin{align} 
\dot{\bm{\psi}} &= -i\left(V+c_0 n + c_1 \bm{F}\cdot\bm{f}\right)\bm{\psi} -i c_2 A_{00} A \bm{\psi}^*.
\end{align}

In order to introduce our full solution to this subsystem, let us first consider the various parts individually.
\subsubsection{Density interaction and trap subsystem}\label{spinsubsys1}
The density interaction term along with the trapping potential causes the system to evolve as
\begin{align}
 \dot{\bm{\psi}} &= -i(V+c_0 n )\bm{\psi}.\label{EqDSS}
\end{align}
This subsystem has $\dot{n} = 0$, thus for a time-independent trap the solution is
\begin{align}
\bm{\psi}(t) &= e^{-i (V+c_0 n) t} \bm{\psi}(0).
\end{align}

\subsubsection{Spin-density interaction subsystem}
The spin-density interaction term causes the system to evolve as
\begin{align}
\dot{\bm{\psi}} &= -ic_1 \bm{F}\cdot\bm{f}\bm{\psi}. \label{EqSDSS}
\end{align}
The commutator of each spin matrix with the evolution operator has the expectation
\begin{align}
\bm{\psi}^\dagger [\bm{F}\cdot\bm{f}, f_\alpha] \bm{\psi} &= i \sum_{\mu\nu} \epsilon_{\alpha \mu \nu} F_\mu F_\nu = 0,
\end{align}
where we have used the standard result (\ref{fcomm}) to derive this.
Thus the subsystem leaves the spin-densities stationary, i.e.
\begin{align}
\dot{F}_\alpha  
&= i c_1 \bm{\psi}^\dagger [\bm{F}\cdot\bm{f}, f_\alpha] \bm{\psi}=0,
\end{align}
so that $\bm{F}\cdot\bm{f}$ is constant in time, and the solution to the subsystem is
\begin{align}
\bm{\psi}(t) &= e^{-i c_1 \bm{F}\cdot\bm{f} t} \bm{\psi}(0).
\end{align}
The evolution operator for this solution takes the general form of a spin-rotation operator, i.e.~$D_{\hat{\mathbf{n}}}(\theta)=e^{-i\theta \hat{\mathbf{n}}\cdot\mathbf{f}}$,  that rotates the state by angle $\theta=c_1Ft$ about a unit vector defined by the spin density $\hat{\mathbf{n}}=\mathbf{F}/F$. Utilizing recent work by Curtright \textit{et al.}~\cite{curtright2014compact} we can analytically write such a spin-rotation as a compact polynomial of the rotation generators ($\mathbf{f}$): 
\begin{align}
e^{-i c_1Ft\, \hat{\mathbf{n}}\cdot\mathbf{f}}
&= I_5 + i \left(\frac{1}{6} \sin (2 c_1 Ft)-\frac{4}{3} \sin (c_1 Ft)\right) \hat{\mathbf{n}}\cdot\mathbf{f} \nonumber \\ &
+\left(\frac{4}{3} \cos (c_1 Ft)-\frac{1}{12} \cos (2 c_1 Ft)-\frac{5}{4}\right) (\hat{\mathbf{n}}\cdot\mathbf{f})^2
\nonumber \\ & 
+i\left(\frac{1}{3} \sin (c_1 Ft)-\frac{1}{6} \sin (2 c_1 Ft)\right) (\hat{\mathbf{n}}\cdot\mathbf{f})^3 \nonumber \\ &
+\left(\frac{1}{4}-\frac{1}{3} \cos (c_1 Ft)+\frac{1}{12} \cos (2 c_1 Ft)\right) (\hat{\mathbf{n}}\cdot\mathbf{f})^4.\label{EqExpF}
\end{align} 
 Thus we can directly compute the exact solution to the spin-density interaction evolution.

\subsubsection{Spin-singlet interaction subsystem}\label{spinsubsys3}
The spin-singlet interaction term causes the system to evolve as
\begin{align}
\dot{\bm{\psi}} &= -ic_2 A_{00}A \bm{\psi}^*,\label{EqSSSS}
\end{align}
which leaves the density  stationary, i.e.
\begin{align}
\dot{n} &= i c_2 A_{00}^* \bm{\psi}^T A \bm{\psi}
- i c_2 A_{00} \bm{\psi}^\dagger A \bm{\psi}^*=0,
\end{align}
however  causes the  spin-singlet amplitude  to evolve as
\begin{align}
\dot{A}_{00} &= -2ic_2 n A_{00}.
\end{align} 
Changing variables as $\bm{\psi} = e^{-i c_2 n t} \tilde{\bm{\psi}}$ leaves the density unchanged, but  rotates out the trivial evolution of the spin-singlet amplitude $A_{00}=e^{-2ic_2nt}\tilde{A}_{00}$, giving
\begin{align}
\dot{\tilde{\bm{\psi}}} 
&= ic_2(n \tilde{\bm{\psi}} - \tilde{A}_{00} A \tilde{\bm{\psi}}^*).
\end{align}
Taking the second time derivative decouples $\tilde{\bm{\psi}}$ from $\tilde{\bm{\psi}}^*$
\begin{align}
\ddot{\tilde{\bm{\psi}}} &= -c_2^2\left(n^2 - |A_{00}|^2\right) \tilde{\bm{\psi}},
\end{align}
with solution  
\begin{align}
\tilde{\bm{\psi}}(t) &= \cos(c_2 S t) \bm{\psi}(0) + \frac{i}{S} \sin(c_2 S t) \left[n \bm{\psi}(0) - A_{00}(0) A \bm{\psi}^*(0) \right], \label{eq:spin-singlet1}
\end{align}
where $S^2 \equiv n^2 - |A_{00}|^2$ and we have used that $\tilde{\bm{\psi}}(0) = \bm{\psi}(0)$. We then rotate back to get the solution to the spin-singlet subsystem,
\begin{align}
\bm{\psi}(t) &= e^{-i c_2 n t} \tilde{\bm{\psi}}(t). \label{eq:spin-singlet2}
\end{align}
 
\subsubsection{Full  potential and nonlinear interaction solution}
We can solve the full nonlinear interaction dynamics if the constituent subsystems (Secs.~\ref{spinsubsys1} to \ref{spinsubsys3}) conserve the interaction terms of each other (i.e.~commute with each other). To justify this we note that for the density and trap subsystem (\ref{EqDSS}), we have $\dot{F}_\alpha=0$  and $\partial_t |A_{00}|^2=0$,  i.e.~the spin-density and spin-singlet interactions are conserved. Similarly, in the spin-density subsystem (\ref{EqSDSS}), we have $\dot{n}=0$  and $\dot{A}_{00}=0$, and for the  spin-singlet subsystem  (\ref{EqSSSS}),  $\dot{n}=0$ and $\dot{F}_\alpha=0$. We note that these relations hold for the general spin-$\mathcal{F}$ case (see Appendix~\ref{app:spinor-identities}).

Thus the subsystems do not affect the dynamics of each other, and the solution to the full nonlinear interaction term is the direct composition of the subsystem solutions:
\begin{widetext}
\begin{align}
\bm{\psi}(t) &= e^{-i c_1 \bm{F}(0)\cdot\bm{f} t}e^{-i[V+(c_0 + c_2) n(0)] t}
\Big\{
\cos(c_2 S t) \bm{\psi}(0) +
\frac{i}{S} \sin(c_2 S t) \left[n(0) \bm{\psi}(0) - A_{00}(0) A \bm{\psi}^*(0) \right] \Big\},\label{EqGenRes}
\end{align}
\end{widetext}
where $e^{-i c_1 \bm{F}(0)\cdot\bm{f} t}$ is to be efficiently evaluated as shown in Eq.~(\ref{EqExpF}).

\section{Numerical Test}\label{SecNumTest}
\subsection{Continuous wave solution}
We test our algorithm numerically using an exact solution to the spin-2 GPE for a translationally invariant system (i.e.~with $V=0$). Since the spinor GPE is  a non-integrable nonlinear partial differential equation it does not have a general solution, and for this test we have developed a continuous wave solution which has the $m=\pm1$ levels unoccupied. 
The form of our solution is
\begin{subequations}\label{cw}
\begin{align}
\psi_{m}&=A_me^{i(k_mx-\omega_mt + \theta_m)},\qquad m=-2,0,2,\label{cw1}\\
\psi_{\pm1}&=0,\label{cw2}
\end{align}
\end{subequations} 
where we determine the relationships between the (positive, real) amplitudes $\{A_m\}$, wave vectors $\{k_m\}$, frequencies $\{\omega_m\}$, and phase shifts $\{\theta_m\}$ for this to be a solution below. 
Taking the stationary limit  ($k_m,\omega_m\to0$) this ansatz approaches the so-called cyclic ground state solution of the uniform system (e.g.~see \cite{Ueda2002a}). The absence of $m=\pm1$ occupation means that our continuous wave solution has zero transverse magnetization, but it will in general have a non-zero $z$-magnetization and spin-singlet amplitude. Thus it is a good candidate to test evolution algorithms for  the spin-2 system.

Since we only have three fields to solve for, we can follow a similar procedure to that used to obtain a spin-1 continuous wave solution  in Ref.~\cite{Tasgal2013}. Requiring Eq.~(\ref{cw})  to be a solution of the  GPE (\ref{spinGPEs})  
we find the following relationship between the quantities determining the phases of the three non-zero fields:
\begin{align}
k_0 &= \frac{1}{2}(k_{2} + k_{-2}), \\
\omega_0 &= \frac{1}{2}(\omega_{2} + \omega_{-2}), \label{eq:wave-freq-rel}\\
\theta_0 &= \frac{1}{2}(\theta_{2} + \theta_{-2} + \bar{n} \pi),
\end{align}
where $\bar{n}$ is an integer.
Similarly, examining the magnitudes of the continuous wave solution in the GPE yields
\begin{align}
\omega_{0} &= \tilde{\omega}_{0} + (-1)^{\bar{n}} c_2 |A_{00}|, \\
\omega_{\pm 2} &= \tilde{\omega}_{\pm 2} + c_2 |A_{00}| \frac{A_{\mp2}}{A_{\pm 2}} + 4q, 
\end{align}
where $\tilde{\omega}_m = \frac{1}{2} k_{m}^2 + c_0 (A_{2}^2 + A_0^2 + A_{-2}^2) + c_1 m (A_{2}^2 - A_{-2}^2)$ and $|A_{00}| = \frac{1}{5}[(-1)^{\bar{n}} A_0^2 + 2 A_2 A_{-2}]$. Asserting Eq.~\eqref{eq:wave-freq-rel}, we get
\begin{align} 
A_0^2 &= (-1)^{\bar{n}+1} 2A_{2} A_{-2} \left(
1 +
\frac{(k_{2} - k_{-2})^2/8 + 4 q}{c_2 (A_{2}-(-1)^{\bar{n}}A_{-2})^2} \right).
\end{align}
Thus we can set $A_{\pm2}$, $k_{\pm2}$, $c_2$, $q$, and $\bar{n}$, subject to the consistency requirement that $A_0$ is real and non-negative, and this determines $A_0$, $k_0$, $\theta_0$, and $\omega_{\pm2,0}$. We note that since the amplitudes $\{A_m\}$ are conserved, the total density, $z$ spin-density and $|A_{00}|$ are all conserved quantities of the solution.

\subsection{Results}
Here we compare the algorithm we have described in this work to two other approaches to solving the spin-2 GPE. In particular, we use both our 
 second-order (S2)  and fourth-order (S4) symplectic algorithms, obtained by composing the exact solutions we have developed using the schemes described in Sec.~\ref{SecSymplectic}. For comparison we have implemented the algorithm developed by Wang \cite{Wang2011a} (W2) which is second-order accurate in time. We also consider a fourth-order accurate method based on the Runge-Kutta method which we denote as (RK4). This type of method is quite commonly used for simulating condensate dynamics and is easily extended to  the spinor case. We note that in implementing this algorithm we have utilized the \textit{interaction picture} technique (e.g.~see \cite{Dennis2013a}) to exponentiate the kinetic energy term and improve the algorithm performance. We observe that for the test problems our implementation of W2 performs about an order of magnitude slower than S2 due to requiring a numerical diagonalization at each mesh point. The other algorithms performance, relative to S2, is similar to the behaviour seen in Ref.~\cite{Symes2016a}.
 
\begin{figure}
	\includegraphics[width=3.1in]{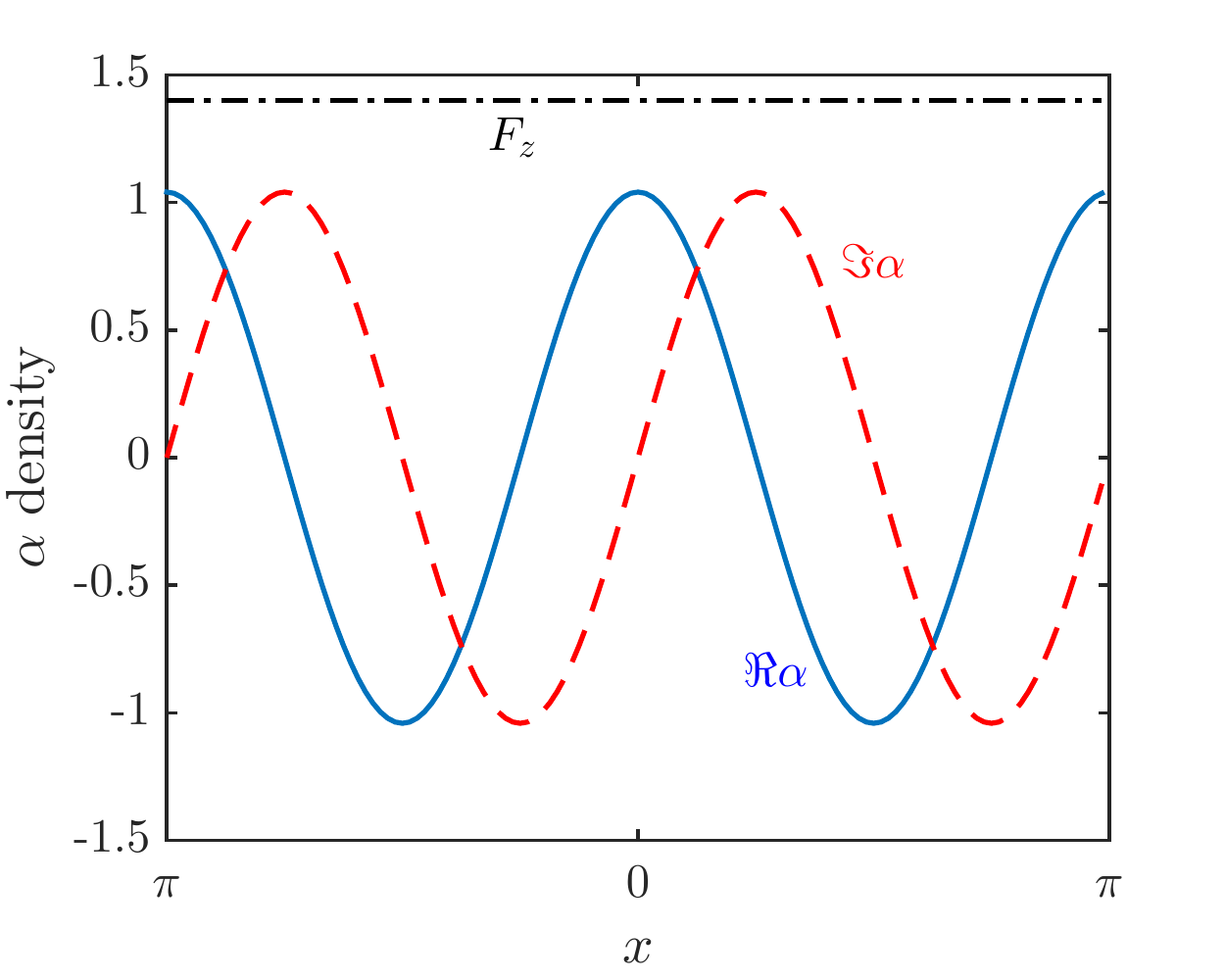}
	\vspace*{-0.5cm}
	\caption{(Color online) The continuous wave solution we use to test our numerics at $t=0$. The  magnetization density (dash-dot line), and the real (solid line) and imaginary (dashed line) parts of the singlet amplitude.
		Parameters of the solution are $c_0 = 10 |c_2|$, $c_1 = 4|c_2|$, $c_2 = -5$, $q = 0$, $\bar{n}=0$, $k_{2} = 3$,   $k_{-2} = -1$, $A_{2} = 0.85$, $A_{-2} = 0.15$.
	}
		\label{fig:psi_cw}
\end{figure}

As a first test we simulate the evolution of a continuous wave solution from $t=0$ to $t=100$ and monitor changes in conserved quantities to quantify algorithm accuracy. 
We assume periodic spatial  boundary conditions, allowing us to implement the kinetic energy operator using Fast-Fourier transforms for all methods.
We have chosen parameters for a continuous wave solution (see Fig.~\ref{fig:psi_cw}) that result in a stable solution with no modulational instability over the time-scales we have considered.
To quantify the changes in conserved quantities  we compute the maximum relative error in conserved quantities    
at each time step, i.e.
\begin{align}
\text{Rel. Error }Q \equiv \left|\frac{Q(t) - Q(0)}{Q(0)}\right|,
\end{align}
where $Q \in \{N, M_z, E\}$ are the three conserved quantities we consider (which we discussed in Sec.~\ref{SecConservQuant}) and $Q(t)$ indicates the quantity evaluated at time $t$ during the propagation.

\begin{figure}
	\includegraphics[width=3.1in]{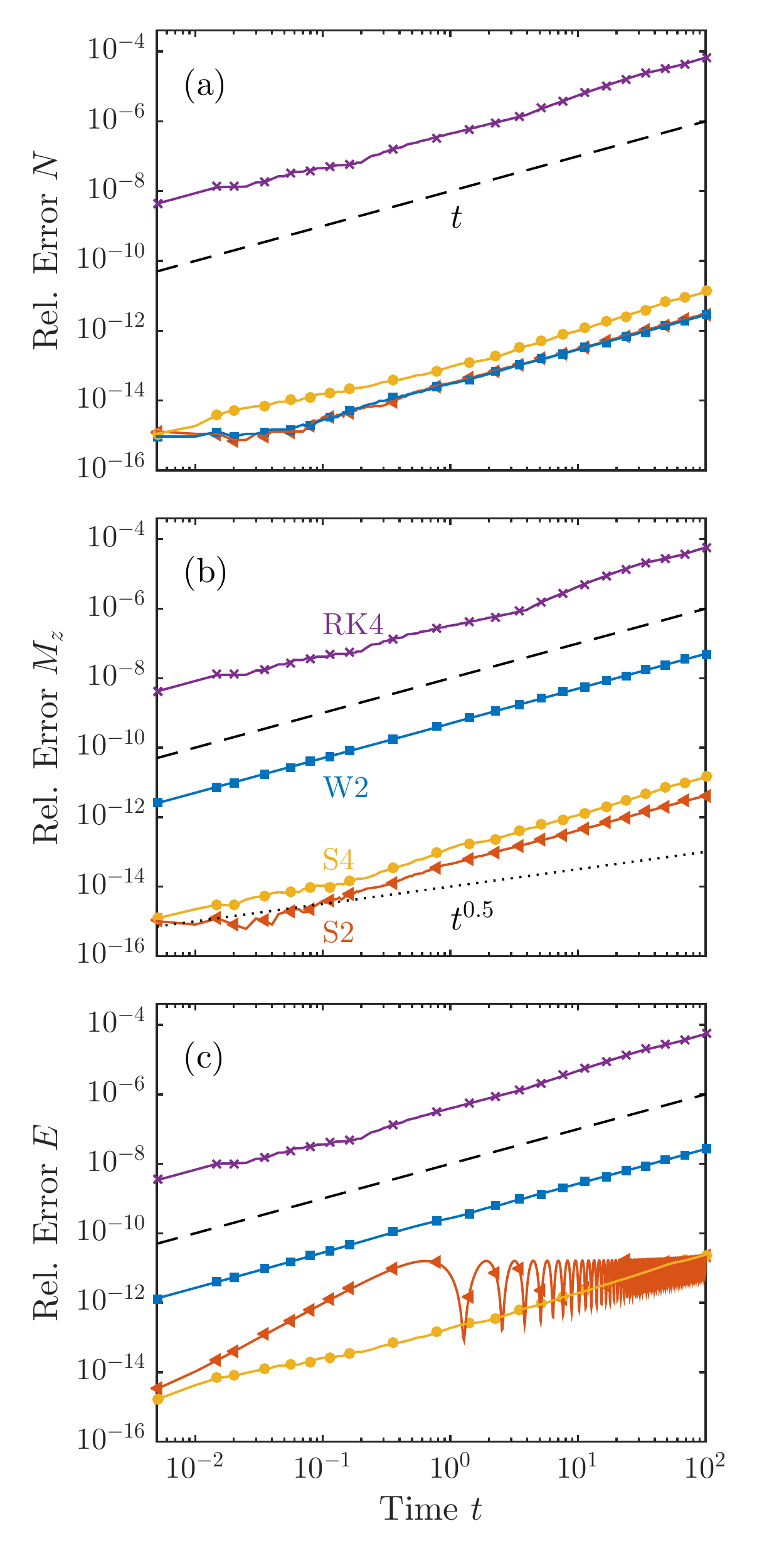}
	\vspace*{-0.5cm}
	\caption{(Color online) Relative errors in the conserved quantities (a) normalization, (b) magnetization and (c) energy for the propagation of the continuous wave solution. The various algorithms, as labeled in (b), all use the same time step ($\tau = 0.005$) and  mesh $\{x_j\}$  of $64$ points over the spatial interval $[-\pi,\pi)$. The initial condition  parameters are given in Fig.~\ref{fig:psi_cw}.  Dashed (dotted) line is a guide to the eye to indicate linear (square root) scaling with time $t$.}\label{FigCWrelE}
\end{figure}

\begin{figure}
	\includegraphics[width=3.1in]{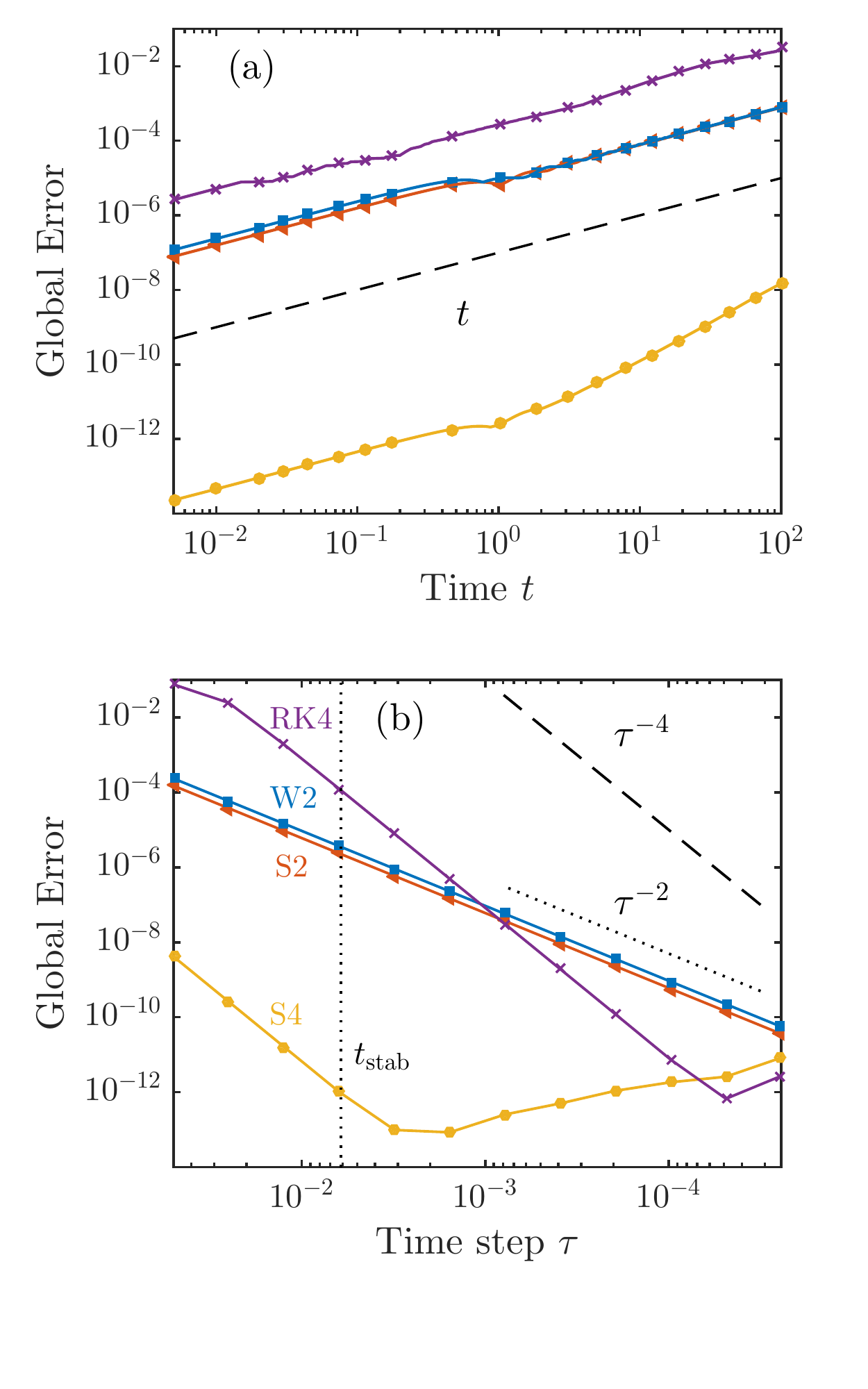}
	\vspace*{-1.2cm}\caption{(Color online) Global error in propagating the continuous wave solution. (a) Global error during a propagation using a time step of $\tau = 0.005$.  Dashed line is a guide to the eye showing linear scaling with time $t$. (b) Scaling of the global error after propagation to time $t_f=0.1$ as a function of the time step $\tau$. The line types for the various algorithms are indicated in (b), as is the estimated stable time step size ($t_{\mathrm{stab}} = 0.006$) due to using a spectral solution for $\hat{f}_A$ \cite{chin2007higher}. The dashed (dotted) lines in (b) indicate fourth-order (second-order) power law scaling with respect to $\tau$. Other parameters for the solution, and mesh, are as indicated in Fig.~\ref{FigCWrelE}. }\label{FigCWglobE}
\end{figure}

Generally the results in Fig.~\ref{FigCWrelE} show that the symplectic algorithms (i.e.~S2 and S4) and the W2 algorithm outperform the RK4 algorithm at conserving the constants of motion. This is expected since symplectic algorithms preserve the geometric properties of phase space.
The energy and  $z$-magnetization are more accurately conserved with the S2 and S4 algorithms than the W2 algorithm. We note that our formulation of the nonlinear subsystem solution Eq.~(\ref{EqGenRes}) exactly conserves $z$-magnetization. Indeed, S2 and S4 initially exhibit $t^{1/2}$ scaling in the $z$-magnetization error, indicating this is limited only by roundoff, until a linear error begins to dominate. The energy error remains bounded for S2 which is a signature of symplectic methods (e.g.~see \cite{mclachlan2006geometric}).
In contrast, the W2 method only approximately treats the nonlinear spin exchange terms so is not symplectic, and this is reflected in the linear growth in error and $z$-magnetization. As a symplectic method, S4 should have bounded energy errors, but here it accumulates a linear error at each time step which is greater than the oscillatory component.

As a second test we can look more closely at the evolution of the spinor field.
Because we know the exact  continuous wave solution we can  quantify the maximum global error at each time step, which we define as
\begin{align}
\text{Global Error } \equiv \text{max}_{x_j, m} \left| {\psi_m^\text{num}(x_j,t) - \psi_m(x_j, t)} \right|,
\end{align} 
where $\psi_m^\text{num}$ denotes the numerically obtained solution, and $\psi_m$ is the exact solution [Eqs.~(\ref{cw})].
In Fig.~\ref{FigCWglobE}(a) we show the growth of the global error over time for a propagation of fixed time step $\tau = 0.05$, i.e.~the same propagation examined in Fig.~\ref{FigCWrelE}. The global error shows that S4 is the best at approximating the exact analytic solution by more than four orders of magnitude, with S2 very similar to W2, and all of these are better than RK4. Global error grows linearly for all methods until $t\sim1$ [i.e.~the time when the energy error starts oscillating  for S2 in Fig.~\ref{FigCWrelE}(c)]. At this point S4 starts having faster than linear error growth, while the other methods continue with linear growth.

In Fig.~\ref{FigCWglobE}(b) we propagate our numerical solution for a fixed amount of time and consider the scaling of the global error with different time steps $\tau$. These results confirm the expected second-order scaling of the global error with $\tau$ for S2 and W2, as well as the expected fourth-order scaling for RK4 and S4.  For this case the prefactor on the global error for RK4 is up to 8 orders of magnitude larger than S4 depending on the choice of $\tau$. With the S4 method the improvement in global error saturates with decreasing $\tau$ at $\tau\sim10^{-3}$, below the stability threshold, while the other methods show improvement in global error down to $\tau < 10^{-4}$. This is because S4 has the most function evaluations per time step and thus accumulates more roundoff errors than the other methods. In our simulations we tried using denser spatial grids and this led to faster accumulation of floating point errors for both composition methods but especially for S4.
We note that due to our choice of a short propagation time, the splitting instability for timesteps greater than $t_\mathrm{stab}$ does not have time to manifest and we observe the expected scaling for all the methods even when $\tau>t_{\mathrm{stab}}$.

\section{Outlook and conclusion}\label{Conclusion}
In this paper we have presented an elegant and highly efficient symplectic algorithm for solving the spin-2 GPE.
We emphasize that our solution for the interaction dynamics (\ref{EqGenRes}) is general, in the sense that it immediately applies to the evolution of all spin-$\mathcal{F}$  systems  (where $\mathcal{F}$ is a positive integer), as long as the evolution  only contains density ($c_0$), spin-density ($c_1$) and spin-singlet ($c_2$) interaction terms [i.e.~evolves according to Eq.~(\ref{spinGPEs})]. Following the procedure we have presented, the density and spin-singlet subsystems are straightforward to compute, with complexity $O(\mathcal{F})$. The spin-density subsystem requires evaluating $2\mathcal{F}$ terms of the compact spin rotation polynomial (using the results of \cite{curtright2014compact}), with a total complexity of $O(\mathcal{F}^2)$. Of course it should be noted that additional spin-interaction terms will occur for $\mathcal{F}\ge3$. There is reason to expect that the treatment of these terms might be challenging (e.g.~see \cite{Ho2000,heinze2007,uchino2008dynamical,ueda2012bose}). Other terms beyond the spinor interactions can also be important. For example, the spin-3 atom $^{52}$Cr has an appreciable magnetic moment causing two such atoms to have a long-ranged dipole-dipole interaction \cite{Santos2006a,Pasquiou2011a}.
 
Our work provides the first symplectic integration schemes for the spin-2 system.
A feature of symplectic methods is that they are useful for obtaining highly accurate solutions of Hamiltonian systems over long time periods.
We have demonstrated that our symplectic algorithms perform well on a specific test case of a continuous wave solution. An important area for future work is to investigate how well these algorithms perform on a wide range of other possible applications.
Particularly promising directions for application of these methods include studies of the long-time coarsening dynamics of the spinor system (e.g.~after being quenched through a quantum phase transition to different kinds of magnetic order, see \cite{Mukerjee2007a,Kudo2013a,Kudo2015a,Williamson2016a,Williamson2016b}), and quantum turbulence \cite{Fujimoto2012a,Fujimoto2012b,Mawson2015a}.

 \section*{Acknowledgments}
 We gratefully acknowledge support from the Marsden Fund of the Royal Society of New Zealand.

\appendix
\section{Spinor identities}
\subsection{Spin matrices}\label{app:spin-matrices}
For spin-2, the spin matrices in the irreducible basis are
\begin{align}
f_+ &= f_-^T = \left(
\begin{array}{ccccc}
0 & 2 & 0 & 0 & 0 \\
0 & 0 & \sqrt{6} & 0 & 0 \\
0 & 0 & 0 & \sqrt{6} & 0 \\
0 & 0 & 0 & 0 & 2 \\
0 & 0 & 0 & 0 & 0 \\
\end{array}
\right), \\
f_z &= \mathrm{diag}\{(2, 1, 0, -1, -2)\}, 
\end{align}
where the in-plane spin matrices are written as
\begin{align}
f_x &= \frac{1}{2}(f_+ + f_-), \\
f_y &= \frac{1}{2i}(f_+ - f_-).
\end{align}
The spin-$\mathcal{F}$ generalization is
\begin{align}
(f_+)_{m,m^\prime} &= \sqrt{(\mathcal{F} + m)(\mathcal{F} - m+1)} \delta_{m-1,m^\prime}, \\
(f_-)_{m, m^\prime} &= \sqrt{(\mathcal{F} - m)(\mathcal{F} + m+1)} \delta_{m+1,m^\prime}, \\
(f_z)_{m,m'} &= m\delta_{m,m'},
\end{align}
where $m,m'\in\{-\mathcal{F},\ldots,\mathcal{F}\}$. These maintain the spin-matrix commutation relations
\begin{align}
[f_a, f_b] &= i \sum_c \epsilon_{abc} f_c, \quad a,b,c \in \{x,y,z\},\label{fcomm}
\end{align}
where $\epsilon_{abc}$ is the completely antisymmetric tensor.
 
\subsection{Interaction commutation relations}\label{app:spinor-identities}
Here we generalize the three subsystems of the spin-2 case to the spin-$\mathcal{F}$ case and show that they commute with each other.
Firstly, the density and trap potential subsystem trivially commutes with all spin-interaction subsystems since its evolution operator is the (scaled) identity.
Next, we generalize the singlet pair matrix to
\begin{align}
A_{m,m'}=\frac{(-1)^{\mathcal{F}-m}}{\sqrt{2\mathcal{F}+1}}\delta_{m,-m'} \equiv \frac{T}{\sqrt{2\mathcal{F}+1}},
\end{align}
where $T$ is the time reversal operator, with the properties
\begin{align}
T^T &= T, \quad T^2 = I, \quad T f_\alpha T = -f_\alpha^*, \\
\bm{\psi}^T T f_\alpha \bm{\psi} &= -\bm{\psi}^T f_\alpha^* T \bm{\psi} = -\bm{\psi}^T T f_\alpha \bm{\psi} = 0,
\end{align}
for $\alpha \in \{x,y,z\}$.
Using this we find that the spin-density subsystem has
\begin{align} 
\dot{\alpha} &\propto \bm{\psi}^T((\mathbf{F}\cdot\mathbf{f}^*) T + T (\mathbf{F}\cdot\mathbf{f})) \bm{\psi} = 0,
\end{align}
and the spin singlet subsystem has
\begin{align} 
\dot{F}_\alpha &\propto A_{00} (\bm{\psi}^T T f_\alpha \bm{\psi})^* - A_{00}^* \bm{\psi}^T T f_\alpha \bm{\psi} = 0.
\end{align}
Thus the three subsystems commute with each other. Using the result in Ref.~\cite{curtright2014compact} for a spin-$\mathcal{F}$ rotation and the generalized form of Eqs.\eqref{eq:spin-singlet1} and \eqref{eq:spin-singlet2}, we can solve the general spin-$\mathcal{F}$ interaction term when it includes terms up to $c_2$.

\end{document}